\documentclass[10pt]{article}
\usepackage{amsmath}
\usepackage{amsthm} %for \theoremstyle{definition}
\usepackage{amssymb,amsfonts,textcomp}
\usepackage{array}
\usepackage{hhline}
\usepackage{graphicx}

\usepackage{tikz}
\usetikzlibrary{trees}
\usepackage{qtree}

\usetikzlibrary{decorations.pathmorphing}
\usetikzlibrary{decorations.markings}
\usepackage{verbatim}

\usepackage[T3,T1]{fontenc} %for \textquotedbl
\usepackage{txfonts} %/pxfonts for \circleright

\usepackage{latexsym} %for \Box
\usepackage{mathrsfs} %for \mathscr

\usepackage{MnSymbol} %for more curved arrow symbol and many others are possible \lcurvearrowright

\usepackage{tikz}
\usetikzlibrary{arrows,shapes,chains,matrix,positioning,scopes}% shapes for circle split
\usetikzlibrary{decorations.pathmorphing}
\usetikzlibrary{decorations.markings}
\usepgflibrary{plotmarks}%for squares and other shapes on graphs lines , 
\usetikzlibrary{patterns}
\usepackage{pdfpages}

\usetikzlibrary{mindmap}%for circle connection bar

\usepackage{graphicx}

\usepackage{multirow}
\usepackage{multicol}

\usepackage{subfig} 

\usepackage{float}
\usepackage{wrapfig}
\restylefloat{figure}
\restylefloat{table}
\usepackage{color, colortbl} %to color rows
\usepackage{float}
\restylefloat{figure}

\newtheoremstyle{slplain}% name
 {.5\baselineskip\@plus.2\baselineskip\@minus.2\baselineskip}% Space above  {3pt}%
  {.5\baselineskip\@plus.2\baselineskip\@minus.2\baselineskip}% Space below
  {\slshape}% Body font {\itshape} %
  {}%Indent amount (empty = no indent, \parindent = para indent)
  {\bfseries}%  Thm head font
  {.}%       Punctuation after thm head
  { }%      Space after thm head: " " = normal interword space;
  {}%       Thm head spec

\theoremstyle{definition}
\theoremstyle{slplain}

\theoremstyle{slplain}

\newcommand{\tfg}{{^\wedge \!}\mbox{{\sf tf}}} %TF-gene pitcher
\newcommand{\tfc}{{^\vee \!}\mbox{{\sf tf}}}% TF-promoter cat
\newcommand{\tfp}{{^\circ \!}\mbox{{\sf tf}}}% TF-protein pitch

\definecolor{currentcolor}{rgb}{0.8 0.4 0.2}%orange currentcolor is to be set to path color and then made lighter

\tikzstyle{stochasticjumpstyle}=[diamond,draw,fill=white,>=latex,>->,dashed]
\tikzstyle{stochasticPathstyle}=[>=latex,>->,dashed]
\tikzstyle{stochasticNodestyle}=[ellipse,inner sep=1pt,text=.,fill=.!20]%[fill=white,inner sep=1pt]%[ellipse,inner sep=1pt,draw,fill=white]
\tikzstyle{blankstyle}=[fill=white,inner sep=1pt]

\def\SnakeSegLen{0.6em}%defines snake segment length for signal jumps  in graphs
\def\SnakeAmp{0.11em}%defines snake amplitude for signal jumps  in graphs
\def\PrePostLen{5mm}

\tikzstyle{sendstyle}=[dashed,line width=1.1pt]%[dotted,ultra thick]

\tikzstyle{splitstyle}=[circle,draw]%not used

\tikzstyle{receivestyle}=[>->,line width=1.1pt,decorate, decoration={zigzag,segment length=\SnakeSegLen, amplitude=\SnakeAmp, pre length=\PrePostLen, post=curveto, post length=\PrePostLen},text=black]

\tikzstyle{receivesigstyle}=[draw,inner sep=2pt,fill=pink!20]
\tikzstyle{receivesigstyle3}=[draw,inner sep=2pt, fill=white]
\tikzstyle{receivesigstyle2}=[ellipse,shade, draw,double,fill=red!10]

\tikzstyle{sendsigstyle}=[diamond,draw,inner sep=1pt, text=black, fill=yellow!80]
\tikzstyle{sendsigstyle3}=[circle,draw, ball color=white]
\tikzstyle{sendsigstyle2}=[diamond,draw,double, inner sep=1pt, fill=white]

\tikzstyle{snakesendstyle}=[*->, decorate, decoration={snake, segment length=\SnakeSegLen, amplitude=\SnakeAmp,  pre length=\PrePostLen, post=curveto, post length=\PrePostLen}]

\tikzstyle{snakesendstyle1}=[line width=1.1pt, decorate, decoration={snake,segment length=\SnakeSegLen, amplitude=\SnakeAmp}]

\tikzstyle{snakesendstyle3}=[decorate, decoration={markings, mark=at position .75 with {\arrow[red,line width=5mm]{>}}, snake, segment length=\SnakeSegLen, amplitude=\SnakeAmp,  pre length=\PrePostLen, post=curveto, post length=\PrePostLen}]

\tikzstyle{snakesendstyle2}=[decorate, decoration={ zigzag,segment length=\SnakeSegLen, amplitude=\SnakeAmp, line around/.style={decoration={pre length=\PrePostLen,post length=\PrePostLen}}}]
\newcounter{foo}
\usepackage[bookmarks=false]{hyperref} %To prevent error in arXiv 

\makeatletter
\AtBeginDocument{\let\nameref\HyPsd@nameref}

\makeatother

\colorlet{anglecolor}{green!50!black}
\definecolor{darkgreen}{rgb}{0 0.6  0}

\definecolor{turquoise}{rgb}{0 0.41 0.41}
\definecolor{rouge}{rgb}{0.79 0.0 0.1}
\definecolor{vert}{rgb}{0.15 0.4 0.1}
\definecolor{mauve}{rgb}{0.6 0.4 0.8}
\definecolor{violet}{rgb}{0.58 0. 0.41}
\definecolor{orange}{rgb}{0.8 0.4 0.2}
\definecolor{bleu}{rgb}{0.39, 0.58, 0.93}

\definecolor{darkross}{rgb}{0.008,0.412,0.471}
\definecolor{middleross}{rgb}{0.012,0.580,0.663}
\definecolor{lightross}{rgb}{0.016,0.749,0.855}
\definecolor{darkblue}{rgb}{0.067,0.008,0.471}
\definecolor{middleblue}{rgb}{0.094,0.012,0.663}
\definecolor{lightblue}{rgb}{0.122,0.016,0.855}
\definecolor{darkpurple}{rgb}{0.471,0.008,0.412}
\definecolor{middlepurple}{rgb}{0.663,0.012,0.580}
\definecolor{lightpurple}{rgb}{0.855,0.016,0.749}
\definecolor{darkbrown}{rgb}{0.471,0.067,0.008}
\definecolor{middlebrown}{rgb}{0.663,0.094,0.012}
\definecolor{lightbrown}{rgb}{0.855,0.122,0.016}
\definecolor{darkolive}{rgb}{0.412,0.471,0.008}
\definecolor{middleolive}{rgb}{0.580,0.663,0.012}
\definecolor{lightolive}{rgb}{0.749,0.855,0.016}
\definecolor{darkgreen}{rgb}{0.008,0.417,0.067}
\definecolor{middlegreen}{rgb}{0.012,0.663,0.094}
\definecolor{lightgreen}{rgb}{0.016,0.855,0.122}
\definecolor{darkocre}{rgb}{0.471,0.298,0.008}
\definecolor{middleocre}{rgb}{0.663,0.420,0.012}
\definecolor{lightocre}{rgb}{0.855,0.541,0.016}

    \definecolor{lightblue}{rgb}{0,0,.7}
    \definecolor{orange}{rgb}{1,.7,0}
    \definecolor{darkorange}{rgb}{1,.4,0}
    \definecolor{darkgreen}{rgb}{0,.5,0}
    \definecolor{darkblue}{rgb}{0,0,.4}
    \definecolor{darkred}{rgb}{.4,0,0}
    \definecolor{gray}{rgb}{.2,.2,.2}
    \definecolor{darkgray}{rgb}{.2,.2,.2}
    \definecolor{shadecolor}{gray}{0.925}
    
\definecolor{darkred}{rgb}{0.65,0,0}
\definecolor{darkblue}{rgb}{0,0,.65}
\definecolor{darkgreen}{rgb}{0,0.5,0}
\definecolor{orange}{rgb}{1,.75,.25}
\definecolor{aqua}{rgb}{0,.25,.75}
\definecolor{grey}{rgb}{.5,.5,.5}
\definecolor{brown}{rgb}{.51,.35,.18}

\definecolor{lightblue}{rgb}{.3,.5,1}
\definecolor{orange}{rgb}{1,.7,0}
\definecolor{darkorange}{rgb}{1,.4,0}
\definecolor{darkgreen}{rgb}{0,.4,0}
\definecolor{darkblue}{rgb}{0,0,.4}
\definecolor{darkred}{rgb}{.56,0,0}
\definecolor{gray}{rgb}{.3,.3,.3}
\definecolor{darkgray}{rgb}{.2,.2,.2}

\definecolor{blue}{rgb}{0,0,1}
\definecolor{red}{rgb}{1,0,0}
\definecolor{pink}{rgb}{.933,0,.933}
\definecolor{green}{rgb}{0.133,0.545,0.133}
\definecolor{shadecolor}{gray}{0.925}

\definecolor{DarkBlue}{rgb}{0.000,0.000,0.545}
\definecolor{DarkChocolate}{rgb}{0.400,0.200,0.000}
\definecolor{DarkCyan}{rgb}{0.000,0.545,0.545}
\definecolor{DarkGoldenrod}{rgb}{0.720,0.525,0.044}
\definecolor{DarkGray}{rgb}{0.664,0.664,0.664}
\definecolor{DarkGreen}{rgb}{0.000,0.392,0.000}
\definecolor{DarkGrey}{rgb}{0.664,0.664,0.664}
\definecolor{DarkKhaki}{rgb}{0.740,0.716,0.420}
\definecolor{DarkLavender}{rgb}{0.400,0.200,0.600}
\definecolor{DarkMagenta}{rgb}{0.545,0.000,0.545}
\definecolor{DarkOliveGreen}{rgb}{0.332,0.420,0.185}
\definecolor{DarkOrange}{rgb}{1.000,0.550,0.000}
\definecolor{DarkOrchid}{rgb}{0.600,0.196,0.800}
\definecolor{DarkPeriwinkle}{rgb}{0.400,0.400,1.000}
\definecolor{DarkPurpleBlue}{rgb}{0.400,0.000,0.800}
\definecolor{DarkRed}{rgb}{0.545,0.000,0.000}
\definecolor{DarkRoyalBlue}{rgb}{0.000,0.200,0.800}
\definecolor{DarkSalmon}{rgb}{0.912,0.590,0.480}
\definecolor{DarkSeaGreen}{rgb}{0.560,0.736,0.560}
\definecolor{DarkSlateBlue}{rgb}{0.284,0.240,0.545}
\definecolor{DarkSlateGray}{rgb}{0.185,0.310,0.310}
\definecolor{DarkSlateGrey}{rgb}{0.185,0.310,0.310}
\definecolor{DarkSmoke}{rgb}{0.920,0.920,0.920}
\definecolor{DarkTurquoise}{rgb}{0.000,0.808,0.820}
\definecolor{DarkViolet}{rgb}{0.580,0.000,0.828}
\definecolor{DeepPink}{rgb}{1.000,0.080,0.576}
\definecolor{DeepSkyBlue}{rgb}{0.000,0.750,1.000}

\tikzstyle{mystyle}=[scale= \PicSize,  %[****Crit. PicSize is not defined*****]
baseline=(current bounding box.center),
nodedescr/.style={fill=white,inner sep=2.5pt},
nodedescr2/.style={draw,circle,fill=white},
description/.style={fill=white,inner sep=2pt},
cancer loop/.style={preaction={draw=white, -,line width=6pt}, line width=1.1pt},
pot1 loop/.style={preaction={draw=white, -,line width=6pt}, red, line width=1.1pt},
pot2 loop/.style={preaction={draw=white, -,line width=6pt}, green, line width=1.1pt},
pot1/.style={preaction={draw=white, -,line width=6pt}, gray, solid, line width=1pt}, 
pot2/.style={preaction={draw=white, -,line width=6pt}, gray, solid, line width=1pt},
inPot/.style={ out=45,in=90,distance=3cm, min distance=2cm, line width=1.1pt, black!75},
inPot1/.style={ out=45,in=120,distance=3cm, min distance=2cm, line width=1.1pt, black!75},
inPot2/.style={ out=-45,in=-120,distance=3cm, min distance=2cm, line width=1.1pt, black!75},
in100Pot1/.style={ out=80,in=90,distance=3cm, min distance=2cm, line width=1.1pt, darkgreen},
in100Pot2/.style={ out=-80,in=-90,distance=3cm, min distance=2cm, line width=1.1pt, brown},
in2Pot1/.style={ out=45,in=90,distance=3cm, min distance=2cm, line width=1.1pt, black!75},
in2Pot2/.style={ out=-65,in=-90,distance=3cm, min distance=2cm, line width=1.1pt, black!75},
selfloop1/.style={ out=45,in=120,distance=6cm, min distance=4cm, line width=1.1pt, red},
selfloop2/.style={ out=-45,in=-120,distance=6cm, min distance=4cm, line width=1.1pt, blue},
loop1red/.style={ out=45,in=120,distance=6cm, min distance=4cm, line width=1.1pt, red},
loop2red/.style={ out=-45,in=-80,distance=6cm, min distance=4cm, line width=1.1pt, red},
cross line/.style={preaction={draw=white, -,	line width=6pt}},
jump1/.style={ out=45,in=135,distance=4cm, min distance=2cm, line width=1.1pt, red, dashed},
jump2/.style={ out=-45,in=-135,distance=4cm, min distance=2cm, line width=1.1pt, blue, dashed},
jumpover/.style={ out=80,in=90,distance=4cm, min distance=3cm, line width=1.1pt, black, dashed},
stochastic jump/.style={>=latex,>->,dashed, line width=1.1pt, green}
]

\def\PicSize{0.5} % 0.5 defines constant PicSize for uniform scale of TikZ pictures

\def\nexttoPicSize2{6.0cm}

\numberwithin{equation}{section}
\usepackage[margin=1.4in]{geometry} %***set page dimensions easily. See geometryPageSize.pdf***

\usepackage[parfill]{parskip}    % Activate to begin paragraphs with an empty line rather than an indent

\begin{document}

\title{\vspace{-7ex}What Transcription Factors Can't Do:\\  \vspace{1ex} \large On the Combinatorial Limits of Gene Regulatory Networks }

\author{Eric Werner \thanks{Balliol Graduate Centre, Oxford Advanced Research Foundation (http://oarf.org), Cellnomica, Inc. (http://cellnomica.com). Thanks: Ulrich Brehm, Francis Hitching.  
\copyright Eric Werner 2013.  All rights reserved. }\\ \\
University of Oxford\\
Department of Physiology, Anatomy and Genetics, \\
and Department of Computer Science, \\
Le Gros Clark Building, 
South Parks Road, 
Oxford OX1 3QX  \\
email:  eric.werner@dpag.ox.ac.uk\\
Website: http://ericwerner.com
}

\date{ } %This is to suppress the printing out of the date.
\maketitle
\thispagestyle{empty}

\begin{center}
\textbf{Abstract}
\begin{quote}
\it \small
A proof is presented that gene regulatory networks (GRNs) based solely on transcription factors cannot control the development of complex multicellular life. GRNs alone cannot explain the evolution of multicellular life in the Cambrian Explosion. Networks are based on addressing systems which are used to construct network links. The more complex the network the greater the number of links and the larger the required address space. It has been assumed that combinations of transcription factors generate a large enough address space to form GRNs that are complex enough to control the development of complex multicellular life. However, it is shown in this article that transcription factors do not have sufficient combinatorial power to serve as the basis of an addressing system for regulatory control of genomes in the development of complex organisms. It is proven that given $n$ transcription factor genes in a genome and address combinations of length $k$ then there are at most $n/k$ k-length transcription factor addresses in the address space. The complexity of embryonic development requires a corresponding complexity of control information in the cell and its genome. Therefore, a different addressing system must exist to form the complex control networks required for complex control systems. It is postulated that a new type of network evolved based on an RNA-DNA addressing system that utilized and subsumed the extant GRNs. These new developmental control networks are called CENES (for Control genes). The evolution of these new higher networks would explain how the Cambrian Explosion was possible. The architecture of these higher level networks may in fact be universal (modulo syntax) in the genomes of all multicellular life.
\end{quote}
\end{center}

\pagebreak\maketitle
\tableofcontents

\section{Introduction}

It is a generally accepted view that networks of protein transcription factors (TFs) control the development of organisms by controlling the expression of genes.  Such networks are known as Gene Regulatory Networks (GRNs).  Here I show that it is impossible for networks based solely on protein transcription factors to control the development of complex organisms.  Hence, transcription factor networks cannot explain the development, origin and evolution of multicellular life.  The reason that transcription factors fail is that they have limited combinatoric capacity.  The address space they can form is too small.   This puts inherent limits on the size of the networks TFs can generate.  In consequence, the addressing system required by complex developmental control networks cannot be based on combinations of protein transcription factors alone.  Furthermore, the protein code is inexact with no apparent canonical, compositional relationship existing between transcription factor combinations and the {\em cis} promoters.  The code of life that is interpreted by the cell to control the development of an embryo is still hidden in the genome.   

An alternative theory of developmental control is presented based on an RNA-DNA addressing system that has the combinatorial capacity to form complex developmental control networks (CENEs) and, thereby, have the capacity to generate complex multicellular life.   CENEs utilize, subsume and control GRNs in order to control gene expression and cell actions.   Given that pre-Cambrian life was controlled by TF-based GRNs then this put inherent limits on the complexity of ancient, pre-Cambrian bacteria and multicellular life.   I propose that there was a switch in the addressing system that made large, complex developmental control networks (CENEs) possible.  These new networks based on RNA-DNA utilized their predecessor TF-networks to control cell action.   The evolution of the combinatorially powerful addressing system and the networks they generated made the Cambrian Explosion possible.  The DNA code of the addressing system and their networks may be universal, modulo syntax, to all multicellular life.   The architectural properties of these networks have direct implications for deciphering the hidden code of life that forms up to 95\% of the noncoding genome of humans and other organisms.  Given, as I have proposed, that stem cell and cancer networks are developmental control networks \cite{Werner2011b}, the discovery of this hidden code has direct relevance to human health.

One of the most fundamental questions of biology is how multicellular life evolved.  What allowed multicellular life to evolve so quickly and with great diversity in the Cambrian Explosion?  What makes complex multicellular life and multicellular development possible?  Here I argue that the current dominant theory of gene regulation cannot explain the development and evolution of complex multicellular organisms.  The development of complex multicellular organisms requires complex developmental control networks (CENEs) that regulate cell actions such as cell division, movement and communication (by subsuming and utilizing GRNs to control gene expression).   These higher level developmental control networks are encoded in the genomes of organisms. Simulations support the general principle that the more complex the organism the more complex its developmental control network.  

\section{Networks are based on addressing systems}
Networks consist of nodes connected by links.  The links in networks are encoded by means of an addressing system that relates one node in a network with another node when those nodes contain addresses that match in some way.   The addresses in an addressing system are created by combining basic address elements (elementary units of combination).  All the possible addresses that can be formed by a given set of basic elements is called the {\em address space} of a given addressing system. 

The implementation of a complex developmental control network requires an addressing system with enough different addresses to form the links between all the different control points or nodes in that network.  Hence, the larger the network, the greater is the number of nodes and links required. Each link requires at least two addresses, one for the source and one for the target. Thus, the larger the network, the larger the address space of possible addresses has to be.   

Therefore, complex multicellular organisms require an addressing system that has sufficient combinatorial power to generate an address space containing all the addresses needed by the developmental control networks that generate such complex multicellular life.  

\section{Gene Regulatory Networks GRNs}
Current theory is that gene regulatory networks (GRNs) based on protein transcription factors (TFs) control the development of organisms.  The links in a GRN network are based on TFs and the promoters they activate by landing and binding to those promoters.  Directed links in a network consist of a source with an address and a target with a matching catching address.  It is thought that the source of a link is a combination of TFs that are caught by {\em cis} promoters of genes.   Thus, the GRN address space encoded in a genome consists of two types of encoded addresses: the source DNA sequences that encode the TF-genes (which generate TF-proteins), and their target {\em cis} promoters which are also encoded as DNA sequences\footnote{The matching relation between the TF-gene address and its {\em cis} promoter address is not straightforward. TFs bind to {\em cis} promoter sites often in combination with other TFs. It has been said there is no TF code or at least not one based on straightforward matching such as antisense molecular binding. }. In order to be potentially activated, TF genes themselves have {\em cis} promoters that are targets for other TFs. Thus, transcription factors genes and TF-proteins and their {\em cis} promoters can form cascades and networks of interlinked nodes. These links of TFs and their {\em cis} promoters form the links in GRNs.   

\section{GRNs control cell actions}
Networks control cell actions.  In bacteria protein transcription factors (TFs) and their promoters control the activation and inhibition of genes.  It has long been assumed that TFs are sufficient to control the development of complex multicellular life \cite{Davidson2002, Davidson2006, Carroll2005, Carroll2008}.   The links in Gene Regulatory Networks (GRNs) are hypothesized to consist of TFs and their promoter targets \cite{Furlong2012}.  It is thought that combinations of TFs have sufficient combinatorial power to form the large address space needed to build complex GRNs.  

I will argue that GRNs based on TFs are fundamentally inadequate as networks for the control of multicellular development because the addressing system on which GRNs are based has limited combinatorial power.  The address space of TFs and their {\em cis} promoters is too limited and hence the networks that can be constructed using this address space are inherently limited and are too small to do the job.  
In other words, GRNs based on TFs cannot generate complex multicellular life because they have limited combinatorial capacity and, therefore, cannot be used as an addressing system for complex developmental control networks (CENEs).  If my proof is correct then there are significant implications for the evolution of multicellular life.  These and their relevance to the Cambrian Explosion are discussed. 

\section{Combinatorics of network addressing systems}
Networks are based on addressing systems. The links in a network relate two nodes by means of addresses that match.  A letter posted in the mail gets to its target because it has the address of the receiver. The address forms a link from sender to receiver. The return address forms a link from receiver to sender.   A system of addresses is usually based on combinations of basic units such as numbers or letters.  For example, the address combination (a,b,c) is different from (a,b,d).  

If the the address elements are ordered where for example, (a,b,c) is different from (b,a,c) then the number of combinations of $n$ elements of length $k$ is $n^{k}$.    
Given the order of the elements does not matter the number of address combinations $C$ of length $k$ that can be generated from $n$ elements is given by the standard formula of combinatorics:
\begin{equation}
C(n, k) =  \left( \begin{array}{c} n \\ k \end{array} \right) =  \frac{n!}{k!(n - k)!}
\end{equation} 

For $n = 1000$ elements and addresses of length $k = 4$, $C(1000, 4) \geq  4.14171247 \times 10^{10}$ address combinations.

Hence, given that there are between 1,000 and 2,600 TF genes in the human genome \cite{Babu2004} one gets a vast number of possible addresses from which (it has been assumed that) complex GRNs can be constructed.  Therefore, TFs appear to have sufficient combinatorial power to serve as the basis for the evolution of the complex control networks necessary for the embryogenesis of complex multicellular life. 

\section{Why TFs won't work as an addressing system}
However tempting, this solution will not work. The problem is that there is an implicit but necessary assumption of combinatorics needed for the generating the address space that fails for protein TFs.   

\subsection{The reusability assumption of combinatorics}
To generate multiple combinations from a basic set of elements it is a {\em central assumption of combinatorics} that these elements can be repeatedly used in the construction of the combinations. For example, the combinations (a,b,c), (a,b,d), (a,b,f) repeat the use of a and b.  If these letters cannot be reused then the address space becomes extremely limited. From $n$ elements for addresses of length $k$ we get at most $n/k$ addresses.  

\subsection{The reusability assumption fails for transcription factors}
Protein transcription factors TFs are usually generated from single copy genes on each homologous chromosome.  To use a TF its gene must be activated by at least one other TF that lands on its promoter.   Each time a TF is used it must be activated by a TF.  That activating TF must in turn be activated by yet a third TF, etc.  One way to get out of this infinite regress is to have a cycle of TFs that activated each other. The simplest cycle is a self-activating TF whose promoter binds its own TF.  Longer cycles lead to a complex cascade of TFs at least one of which loops back to the start TF to initiate the same cascade over and over again.  In fact, this actually happens when maternal TFs, which are inherited from one or both parents and located in the fertilized egg, bootstrap and initiate embryogenesis.  

The problem is that the maximum length of such a cascade is equal to the total number TF-genes encoded in the genome.  If the network is not just a cycle but is instead a complex developmental network such as a tree with loops, then the maximum number of TFs available as address links between control nodes for use in the network is again limited to the copy number of TF-genes in the genome.  If the minimum address is of length $k$ then with $n$ TFs there are only $n/k$ addresses available.  For 1000 TFs and addresses of length 2, we get only 500 nodes in the network. 

This puts severe restrictions on the possible complexity of such control networks.   The reason is that the address space is too limited because of the failure of the reusability assumption of combinatorics. Each use of a TF requires a new, separate TF.  A hierarchy of TFs will not help ameliorate this fatal flaw.  Because, each reuse of a higher level TF2 to activate a lower level TF1 requires a new TF3 to activate TF2 which then activates TF1. 

\section{Why {\em cis} regulatory sites are not sufficient}
One might think to avoid this limitation by constructing {\em cis} regulatory promoters in arbitrary combinations\footnote{This has been proposed by Carroll when he assumes all evolutionary change occurs in {\em cis} promoter regions \cite{Carroll2005,Carroll2008}.}.  Clearly, the binding sites of protein TFs can be and are repeated in genomes. Hence, we can construct arbitrarily many {\em cis} promoter addresses that catch TFs.  Thus the possible {\em cis} address space does satisfy the formula $C(n,k) = \mbox{\sf CisAdr}(n, k) = n!/k!(n - k)! $  giving us a vast space of possible addresses. 

The problem with this counter move is that to control the unique activation of these addressable {\em cis} areas of genes, we require unique combinations of TFs.  But we only have at most $n/k$ such TF address combinations available.  Hence, we either have entropic TF addressing or the TF addresses are too limited to cover the {\em cis} address space.  

\subsection{Notation and definitions}
To see why, we need some notation and definitions.  A given {\em cis} address of length $k$ has the form $\tfc_{1}$,$\tfc_{2}$, \ldots , $\tfc_{k}$  where each $\tfc_{i}$ is the catching address of some TF-protein $\tfp_{i}$.  (When clarity requires it we use the notation $\tfg$ for the TF-gene that generates the TF-protein, otherwise, if it is clear from the context TF will refer either to the TF-gene that generates the TF-protein or to the TF-protein itself.)  We say an address has {\em OR-address matching} ({\em OR-addressing}) if the gene it controls is activated if any one of the catcher's $\tfc$s is {\em loaded} by its matching $\tfp$ protein.  With OR-addressing only one of the TFs in a TF address combination has to match a catching {\em cis} promoter address to activate the gene.  One has {\em AND-address matching} ({\em AND-addressing}) if all of the catcher $\tfc$s have to be loaded by their matching $\tfp$ to activate the gene that the {\em cis} address controls.  With AND-addressing each TF-protein  in a TF-protein address combination has to match a corresponding catching subaddress in a {\em cis} promoter address.

Let AdrTF$(n,k)$ be the space all possible addresses of length $k$ that can be formed from $n$ catching $\tfc$s.  Given the combinations are unordered, the number of elements in AdrTF$(n,k)$ is $C(n,k)$.  If, on the one hand, we have OR-address matching then there is an inherent address entropy (ambiguity) such that any given $\tfg$ combination matches (and activates) a potentially large subset of {\em cis} promoter addresses $\tfc$ in AdrTF$(n,k)$.  

\section{Control entropy}
While OR-addressing is well and good for activating sets of genes (which are indeed needed for some cellular actions or processes), {\em cis} addressing combinatorics using only $\tfg$s and their catcher $\tfc$s, cannot be used for fine grained, complex global control of development because address entropy leads to network entropy and network control entropy.  It is like trying to steer two cars and drive them to two different cities having only one steering wheel.  When you get to the corner one car has to go the left and other has to go the right.  This is control entropy. 

If, on the other hand, we have AND-addressing then a given TF address combination matches only one {\em cis} address in AdrTF$(n,k)$.    AND-addressing can only pick out a singleton subset of addresses out of the range of possible {\em cis} addresses in AdrTF$(n,k)$ .   Hence, AND-addressing cannot be used to cover the possible {\em cis} promoter addressing space. Hence, it cannot be used to build complex developmental control networks.  Therefore, either TF-addressing is either too general or too restrictive.  

The set of {\em cis} addresses actually used as promoters to genes need not be and in general will not be equal to the set of all possible addresses AdrTF$(n,k)$.  In a genome a {\em cis} catching address can be repeated over and over again.  Only the TF-genes that generate TF-proteins that land on and load {\em cis} catching promoter addresses have limited copy numbers.  Given any TF-gene $\tfg$, then for any gene in a genome we could construct a {\em cis} promoter address $\tfc$ for that gene so that the $\tfg$ address matches the {\em cis} address.  Hence for any gene and any $\tfg$ we can construct a link from the $\tfg$ to the gene by placing a catching {\em cis} promoter address $\tfc$ in front of the gene.  However, because of the limited copy number of  TF-genes ($\tfg$s) relative to the number of genes, one cannot have a one-to-one mapping from $\tfg$s to genes. Therefore, a given $\tfg$ either activates one gene or a set of genes.  Given $n$ $\tfg$s they can at most control $n$ unique genes or $n$ subsets of genes.   Therefore, while we can repeat a whole {\em cis} address, this would again lead to address entropy, which leads to network entropy and control entropy. 

\subsection{Example of control entropy}
The situation is analogous to trying to control a candy vending machine that has an array of  5 by 10 candy box selections with only two buttons, but where button combinations are not allowed. The engineer faced with controlling this machine either has to associate a button with a unique candy box or the engineer has to associate a button with a subset of boxes. Then depending on the how the engineer programmed the button if you push one button you either get one candy or a set of candies.  Without combinatoric addresses you cannot program this machine to give the user the choice of any particular candy.  So it is with the use of TFs in genomes to select unique genes.  

Given 8 TF-genes for creating links to the next nodes and given no repeats are allowed, we can get at most a linear cascade of 8 TF-genes connecting 9 nodes, or we can get a balanced binary tree of  depth $\leq \log_{2}(8) = 3$, or we can get something in between.  Still the nodes in this example network are limited to 9 and the links are limited to 8. 

\section{Complex development requires complex networks}
In embryogenesis, and more generally the development of multicellular organisms from a single cell, there has to be temporal and spatial control of not just genes but processes and cell actions where genes are used over and over again in different contexts in space and time.  Since $n$ copies of TFs can control at most $n$ nodes in a network, they are inadequate as a basis of control for the evolution and development of complex multicellular life. 

Simulations show that as the complexity of multicellular organisms increases there is to be a corresponding increase in complexity of the control networks that generate such organisms. To control processes that are repeatedly used at various points in development requires repeated use of control nodes.  But this repeated use must itself be controlled by yet another level of control. TFs alone, as we have shown, cannot be the source of this higher level of control.   

\subsection{TFs cannot generate complex networks}
This result shows that TFs cannot be the source of complexity in the evolution of multicellular organisms.  Instead an alternative addressing system must form the foundation for the complex developmental control networks necessary for complex multicellular life.   One possibility is an addressing system based on RNA or DNA. Both RNA and DNA sequences form ordered addresses that can match a corresponding DNA addresses either directly or antisense.  Given the 4 base pairs as the combinatory elements of an address of length k, we get $4^k$ possible addresses.  Unlike TFs, the combinatory elements (T,  C, A, G for DNA and U, C, A, G for RNA) are repeatable. For short DNA or RNA sequences of length 20 this generates a very large address space ($4^{20}$ or over one trillion addresses).  Another advantage of this RNA-DNA addressing system is that, in addition to repeatability of the combinatoric elements that make up addresses, whole addresses can also be repeated over and over again.  The result is a highly flexible addressing system.  

\subsection{Evolution of the addressing system antedates the evolution of complex networks}
Therefore, the evolution of complex multicellular life required a switch in the addressing system from protein based TF-addressing to RNA-DNA addressing.  It may have been that ancient primitive, simple multicellular organisms were controlled by networks based on a protein TF addressing system.  Since the complexity developmental networks based on a TF addressing system have inherent limits, the organisms they generate would have a correspondingly limited complexity in function and morphology and possibly have redundant, repeated substructures.  Their multicellular structures would be highly redundant if based on cascades of interlinked TFs that formed cycles of control.  And, indeed we see this in pre-Cambrian stromatolites and linear growth structures.  So too, we see it in the fern like repetitive organisms immediately before the Cambrian Explosion.  
% ****refs and names***.

\subsection{The Cambrian Explosion}
The Cambrian Explosion was made possible by the evolution of an addressing system that permitted the formation of complex developmental networks. The nodes of these networks controlled cell actions and processes by utilizing and controlling more ancient TF based control networks that regulated the expression of genes in the cell.  The agents of cell actions are ultimately constructed using proteins produced by the coordinated expression protein coding genes.  Hence, the more ancient genetic control by transcription factors was subsumed by higher-level control networks that were based on a more powerful addressing system.  

\subsection{A subsumption architecture of genome-cell control}
Thus, the eukaryotic cell that forms the basis of multicellular life has a control architecture similar to the robotic subsumption architecture proposed by Brooks \cite{Brooks1986}.  

\begin{figure}[H]
\centering
{
\includegraphics[scale=0.5]{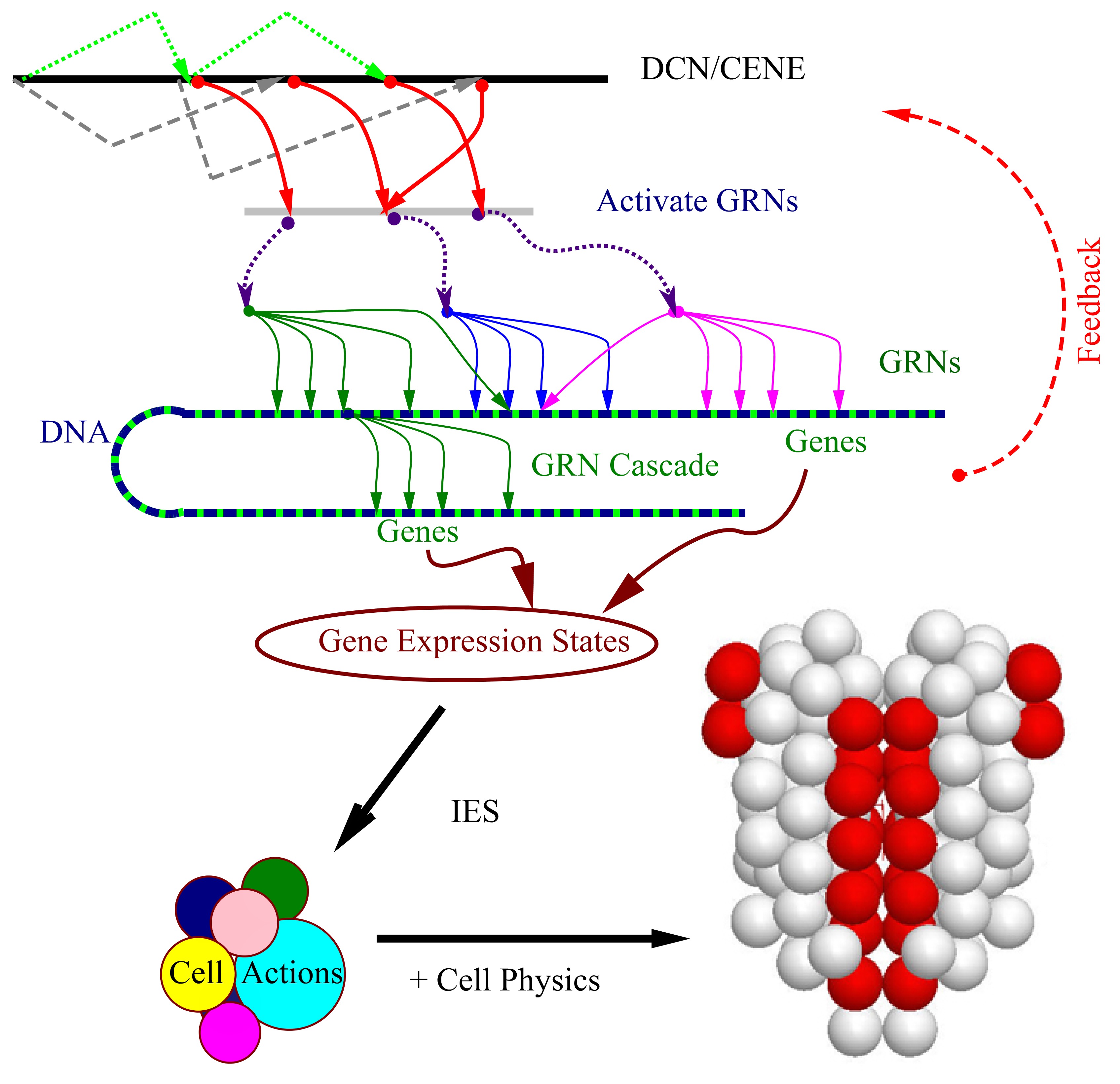} 
}
\caption{A subsumption architecture for CENE-GRN-Cell control in embryonic development}
\label{fig:GenomeArchitecture}
\end{figure}

Therefore, gene regulatory networks (GRNs) based on TFs were subsumed by higher-level developmental control networks (CENEs).  I have called developmental control networks CENEs for control genes, since these networks play the dominant role in determining the phenotype of multicellular organisms.  Thus, these CENEs or developmental control networks are more like what Mendel meant when he discovered the units of inheritance \cite{Werner2011a}. 

\subsection{The separation of GRNs from developmental control networks made evolution possible}
The division of developmental control from genetic control separated the local control of survival of the individual cell from the global control of multicellular development.  This permitted the autonomous evolution multicellular morphology and function from the evolution of the genetic code that controlled local cell actions, tactics and survival.  Once simple developmental control networks were formed based on the new addressing system, they could be duplicated and modified much like genes were duplicated and modified \cite{Ohno1972,Ohno1972a}.   This switch in the addressing system made complex developmental control networks possible.  The rapid evolution of body forms in the Cambrian Explosion was based on network duplication and modification by network transformations such as network link switching and node subsumption transformations that led to different directives to the genetically controlled cells by altering gene expression states.  

With the new addressing system both the source address and the target address could be repeated. And they could change. Hence, the disadvantages of a TF-based addressing system and the resulting control entropy were overcome. 

CENEs are interpreted and executed by the cell. I have called this the Interpretive-Executive System or IES.  While the network links control the temporal order of directives to the developing cells, the network nodes link to lower level GRNs that control cell actions (See figure \ref{fig:GenomeArchitecture}).  

\section{GRNs are not sufficient for complex bacterial control}
Real GRNs in bacteria may be purely TF based networks but they may also be in part controlled by a richer addressing system that involves RNA \cite{Shapiro2009,Shapiro2011}.  
In fact, beyond a certain point, as the control strategies of bacteria became more complex, the TF-based addressing system and pure TF-based control networks would become inadequate to reflect and generate such control tactics and strategies.   Hence, even prokaryotes beyond a given complexity bound will have had to evolve and utilize a more combinatorially adequate addressing system based on RNA or DNA.  Indeed, this encoding of cell action control may have been the evolutionary predecessor of the encoding of cell action control required for complex multicellular development.  The overall point is that these complexity limits of TF-based control networks would apply even to bacterial control. 

While the origin of complex multicellular life in the Cambrian Explosion required a switch in the addressing system from transcription factor control to an addressing system with capacity to match the complexity of the evolving multicellular organisms, the actual evolution may have been more gradual converting bacterial RNA control into a higher level RNA based developmental control network.  

\section{Cell physics}
The physics of the cell plays a complimentary role in generating the outcome of the interpretation and execution of developmental control networks (see \cite{Newman2009}).  However, physics by itself, without control by the genome, leads to structures akin to mineral and crystal formation, which are highly redundant structures often with random global and local organization.  This is because physics does not contain the global control information to determine complex morphology.  This difference of physically formed structures and structures formed by developmental control networks allows us to distinguish fossils from mineral deposits. 

\section{Conclusion}
In summary, the complexity of organisms requires a corresponding complexity in the developmental control networks in the genome that direct embryogenesis.  Since the links that make up networks are implemented by means of an addressing system, complex networks require a sufficiently large address space.  Protein transcription factors (TFs) fail to provide an adequate addressing system because they do not satisfy a fundamental assumption of combinatorics, namely their repeatability. Hence, transcription factors cannot form the combinations required to generate a large address space.  Therefore, the ancient primitive addressing system used by pre-Cambrian bacteria had to be supplemented by a new combinatorially more powerful addressing system based on RNA-DNA for the development of complex multicellular organisms to become possible.  

The developmental control networks based on the richer addressing systems I call CENEs (for control genes).  This new system is linked in to the older TF-based addressing system that controlled gene expression.   Thus, gene regulatory networks (GRNs) were subsumed by higher-level CENE networks allowing CENE networks to control cell actions by controlling gene expression via GRNs.   

The relative autonomy of developmental control networks (CENEs) from the lower level cell control exercised by the TF-based gene regulatory networks (GRNs) allowed the autonomous evolution of CENE networks without affecting the survivability of the cell.   Rapid evolution of body forms became possible precisely because increase in complexity of CENE networks did not require a corresponding increase in complexity of the GRNs or in the controlled genes.   

While cell physics plays an important role in development it is not the primary driver of diversity in evolution, since much of cell physics is common to all multicellular organisms.  The separation of control by higher-level CENE networks from lower level TF-based networks, also explains the limited evolution of TFs and other genes shared by most metazoans.  Thus humans share most of their genes with chimpanzees, mice, flies and worms.   

The proposed architecture of networks and the addressing systems controlling the development of multicellular life also has consequences for decoding the so far hidden code in the non-protein coding genome that makes up 95\% of the human genome. The code may be universal, with slight variations, in all multicellular life.  Once this higher level control network code is deciphered, diseases such as cancer controlled by such networks will potentially be curable by transforming cell cancer networks back into harmless networks.  So too, full control of nerve, tissue and organ regeneration will become possible based on a new understanding of stem cell networks \cite{Werner2011b}.  Thus, these results imply that we need a major paradigm shift from a gene-centered paradigm to a developmental control network CENE-centered paradigm in order to understand multicellular diseases, development, and evolution.  

\footnotesize %\small %\footnotesize %\tiny
\bibliographystyle{abbrv}
\bibliography{WernerTFlimitsInformal}

\end{document}